\documentclass[12pt] {article}

 \font\gotb eufm10 scaled \magstep1
\newcommand{\bb}{\bibitem}
\newcommand{\cc}{\cite}
\newcommand{\vp}{\varphi}
\newcommand{\sss}{\sigma}
\newcommand{\al}{\alpha}
\newcommand{\Om}{\Omega}
\newcommand{\om}{\omega}
\newcommand{\lt}{\left}
\newcommand{\rt}{\right}
\newcommand{\lll}{\lambda}
\newcommand{\F}{{\cal F}}
\newcommand{\Ss}{\hat S^2_}
\newcommand{\RR}{\bar R}

\newcommand{\am}{\hat a^-}
\newcommand{\ap}{\hat a^+}
\newcommand{\R}{\hat R}
\newcommand{\s}{\hat S}
\newcommand{\K}{\hat K}
\newcommand{\I}{\hat I}
\newcommand{\Q}{\hat Q}
\newcommand{\E}{\hat E}

\newcommand{\A}{\hat A}
\newcommand{\B}{\hat B}

\newcommand{\p}{\hat p}
\newcommand{\PP}{\hat P}
\newcommand{\po}{\Psi_0}
\newcommand{\pst}{\Psi_\tau}
\newcommand{\pht}{\Phi_\tau}
\newcommand{\pt}{\p_\tau}
\newcommand{\aA}{\tilde A}
\newcommand{\HH}{\hat H}
\newcommand{\tvp}{\tilde \vp}
\newcommand{\AAA}{\hbox{\gotb A}}
\newcommand{\AC}{$\hbox{\gotb A}_{cl}$}
\newcommand{\QQ}{\hbox{\gotb Q}}
\newcommand {\QQQ}{\hbox {\gotb Q}_{\xi}}
\newcommand {\qqq}{\hbox {\gotb Q}_{\xi'}}
\newcommand {\vx}{\vp_{\xi}}
\newcommand {\vpx}{\vx(\A)}
\newcommand{\BB}{\hbox{\gotb B}}
\newcommand{\bea}{\begin{eqnarray} \label}
\newcommand{\eeq}{\end{equation}}
\newcommand{\beq}{\begin{equation} \label}
\newcommand{\eea}{\end{eqnarray}}
\newcommand{\nn}{\\ \nonumber}
\newcommand{\rr}[1]{(\ref{#1})}

 \author{D.A.Slavnov}

\title{Postulates of quantum mechanics and phenomenology}

   \date{}

\begin{document}

  \maketitle
    \begin{center} {\it  Department of Physics, Moscow State
University,\\  Moscow 119899, Russia. E-mail:
slavnov@goa.bog.msu.ru }
 \end{center}

 \begin{abstract}

We describe a system of axioms that, on one hand, is sufficient
for constructing the standard mathematical formalism of quantum
mechanics and, on the other hand, is necessary from the
phenomenological standpoint. In the proposed scheme, the Hilbert
space and linear operators are only secondary structures of the
theory, while the primary structures are the elements of a
noncommutative algebra (observables) and the functionals on this
algebra, associated with the results of a single observation.
\end{abstract}

\section {Introduction}

Hilbert space and the linear operators in it are the basic
concepts in modern quantum mechanics. Von Neumann~\cc{von} gave a
transparent mathematical formulation of quantum mechanics based on
these concepts. The Hilbert space formalism became the
mathematical basis of the immense achievements in quantum
mechanics. But such a seemingly perfect construction of the theory
is not free from shortcomings, which was noted by~\cc{seg}: "These
axioms are technically simple but intuitively, they are completely
unclear and seem to be ad hoc".

The same can be expressed in other words as follows.  The axioms
of the standard quantum mechanics are phenomenologically
sufficient, i.e., we can describe practically all observable
effects on the basis of these axioms. At the same time,  their
phenomenological necessity is not clear.  Such a situation may
lead to undesirable consequences.  A theory constructed based on
these axioms may prove to be "overdetermined". Numerous paradoxes
such as the Einstein-–Podolsky-–Rosen paradox~\cc{epr}, the
"Schr\"odinger cat"~\cc{sch,hom},  and so on are indirect evidence
supporting the existence of such a danger in the standard quantum
mechanics. We mention at once that there are different opinions
about the paradoxes.  The most active adherents of the standard
quantum mechanics deny the existence of any paradoxes.

In any case,  it is extremely desirable that only statements that
can be directly verified in an experiment are chosen as
postulates.  Otherwise,  when remote consequences are verified,
there always exists a danger that we have not verified all of
them,  and the unverified ones may contradict the experiment.
Although such a system of axioms may be less effective
technically,  it will be more reliable.  In this article,  we try
to give such a formulation of the main postulates of quantum
mechanics.  The proposed system of axioms is not the ideal one in
this respect; we only want to minimize the technical part.

Another disadvantage of the approach based on using the Hilbert
space is that the Hilbert space is foreign to classical physics.
In classical physics,  the phase space is used,  which in turn is
foreign to the standard quantum mechanics.  A philosopher would
say that classical and quantum physics use different paradigms.
This term can be understood as some rules of the game that are
given a priori.  It would be extremely desirable to construct a
scheme covering both classical and quantum mechanics.  In this
article,  a corresponding attempt is made.  This by no means
implies that quantum mechanics is reduced to classical mechanics.
At the same time, the construction is performed in the framework
of the classical paradigm.

The classical paradigm is primarily the classical formal logic and
the idea about the existence of a causal connection between both
physical phenomena and logical statements.  Secondarily, it is the
assumption about the existence of physical realities that are the
bearers of the causes of physical phenomena and the assumption
that probabilistic statements satisfy the classical Kolmogorov
probability theory.

It is commonly assumed that these postulates are incompatible with
the mathematical scheme adopted in quantum mechanics.  Here,  we
try to prove the contrary.  This article is a further development
of the ideas formulated in~\cc{slav1,slav2}. The contents of this
article corresponds to paper~\cc{slav3}.

\section{Observables,  states,  and measurements}

We begin by considering a classical physical system.  For such a
system,  "observable" is a basic notion. This notion seems
self-evident,  requiring no exact definition.  Heuristically,  an
observable is an attribute of the studied physical system that can
be given a numerical value by means of a measurement. In what
follows,  we assume that the measurement is ideal, i.e.,  is
performed with ideal accuracy.

The main property of observables is that they can be multiplied by
real numbers,  added,  and multiplied by each other.  In other
words,  they form a real algebra~\AC.  To define a physical
system, we must establish the relations between different
observables, i.e.,  to fix the algebra~\AC.  In the classical
case, the algebra turns out to be commutative.

Fixing an observable $\A$ still tells nothing about the value A
that will be obtained as a result of a measurement in a concrete
situation.  Fixing the values of observables is realized by fixing
the state of a physical system.  In mathematical terms,  this
corresponds to fixing a functional $\vp(\A)$ on the algebra~\AC.

We know from experience that the sum and product of observables
correspond to the sum and product of the measurement results:
$$
 \A_1+\A_2 \to A_1+A_2, \qquad \A_1\A_2 \to A_1A_2.
$$

In this connection,  the following definition~\cc{rud} is useful
in what follows.

\

{\sc Definition 1.} {\it   Let \BB{} be a real commutative algebra
and $\tvp$  be a linear functional on \BB{} If
$\tvp(\B_1\B_2)=\tvp(\B_1)\tvp(\B_2) \mbox{ for all }\B_1\in\BB
\mbox { and } \B_2\in\BB $,  then the functional $\tvp$ is called
a real homomorphism on the algebra \BB{}.}

\

Using this definition,  we can say that a state is a real
homomorphism on the algebra of observables. Physically,  a state
is some attribute of the physical system separated from the
environment.  A system can be either isolated or nonisolated.  In
the latter case,  we assume that the external influence is reduced
to the action of external fields.  We suppose that a state is
determined by some local reality including the internal  structure
of the physical system as well as the structure of the external
field in the localization domain of the system.

In the measurement process,  the physical system is influenced by
the measuring instrument.  According to the character of this
influence,  measurements can be divided into two types:
reproducible and nonreproducible.  The characteristic feature of
reproducible measurements is that a repeated measurement of an
observable gives the same result in spite of the perturbation
suffered by the system in each measurement.  We assume that the
system is isolated from external influences in the interval
between measurements and that we are able to take a possible
change of the observable in the process of free evolution into
account.  In what follows,  we are mainly interested in
reproducible measurements.  We therefore include the
reproducibility requirement,  if the contrary is not explicitly
mentioned,  in the term "measurement".

The reproducibility problem is particularly interesting if the
measurements of several observables for the same physical system
are performed.  For example,  we suppose that we first measure an
observable $\A$,  then an observable $\B$,  then again the
observable $\A$,  and finally the observable $\B$.  If the results
of the repeated measurements coincide with the results of the
initial measurements for each observable,  then we say that such
measurements are compatible.  If there exist devices for
compatible measurements of $\A$ and $\B$,  then these observables
are said to be compatible or simultaneously measurable.

We know from experience that all observables are compatible for
classical physical systems.  In contrast,  there are both
compatible and incompatible observables in the quantum case.  In
the standard quantum mechanics,  this fact is qualified as the
"complementarity principle".  We regard it simply as evidence of
the fact that measuring two incompatible observables requires
mutually exclusive instruments.

Although the observables also have algebraic properties in the
quantum case, it is impossible to construct a closed algebra from
them that would be real,  commutative,  and associative.  On the
other hand, it is extremely desirable from the technical
standpoint to have a possibility to use the advanced mathematical
apparatus of the theory of algebras.  We therefore adopt the
following compromise variant of the first postulate.

\

 {\bf Postulate 1.}

{\it   The observables form a set $\AAA_+$ of Hermitian elements
of an involutive,  associative,  and (generally) noncommutative
algebra~\AAA{} satisfying the conditions that for every element
$\R\in\AAA$  there is a Hermitian element $\A$ $(\A^*=\A)$ such
that  $\R^*\R=\A^2$  and that if $\R^*\R=0$, then $\R=0$.}

\

We assume that the algebra has a unit element~$\I$.  The elements
of the algebra~\AAA{} are called dynamical quantities.

Postulate 1 is not completely free from a technical component,
but the latter is essentially smaller than in the corresponding
postulate of the standard quantum mechanics.  There,  it is
additionally postulated that dynamical quantities are linear
operators in a Hilbert space.  This fact can be hardly considered
self-evident.

The following postulate is an immediate consequence of quantum
measurements.

\newpage

{\bf  Postulate 2.}

{\it  Compatible (simultaneously measurable) observables
correspond to mutually commuting elements of the set $\AAA_+$}.

\

In connection with this postulate,  real commutative subalgebras
of the algebra~\AAA{} are especially important for us.  Let $\QQ
\quad (\QQ\equiv \{\Q\}\subset\AAA_+)$ denote a maximal real
commutative subalgebra of the algebra~\AAA{}. This is the
subalgebra of compatible observables.  If the algebra~\AAA{} is
commutative (an algebra of classical dynamical quantities), then
such a subalgebra is unique.  If the algebra~\AAA{} is
noncommutative (an algebra of quantum dynamical quantities), then
there are many different subalgebras $\QQQ$ $(\xi \in \Xi)$.
Moreover, the set $\Xi$  has the cardinality of the continuum.
Indeed, even if the algebra~\AAA{} has two noncommtative Hermitian
generators $ \A_1 $ and $ \A_2 $, then every real algebra
$\QQ_{\al} $ with the generator $ \B (\al) = \A_1\cos\al +
\A_2\sin\al $ is an algebra of type~\QQ.

In both the classical and quantum cases,  the value of a variable
is revealed in an experiment.  In other words,  the result of a
measurement is described by some functional whose domain is the
set $\AAA_+$.  But the structure of this functional turns out to
be much more complicated in the quantum case than in the classical
case.

We can separate a subsystem described by the subalgebra $\QQQ$
(the corresponding dynamical variables) from the quantum physical
system described by the algebra \AAA{}.  Because the subalgebra
$\QQQ$ is commutative,  the separated subsystem can be considered
classical.  Its state can be described by a functional $\vx$  that
is a real homomorphism on the algebra $\QQQ$.  Of course, this
classical subsystem is not isolated from the rest of the quantum
system,  but isolatedness is not a necessary condition for
separating a classical subsystem. A state of a classical system is
determined by some local physical reality; it would be strange if
the same were not true for the classical subsystem.

\

{\sc  Definition 2.}  {\it  A multilayer functional $\vp=\{\vx\}$
is the totality of  functionals $\vx$,  where $\xi$ ranges the set
$\Xi$ and the domains of the functionals $\vx$ are the subalgebras
$\QQQ$.}

\

In quantum measurements in each individual experiment,  we deal
only with observables belonging to one of the subalgebras $\QQQ$.
The result of such a measurement is determined by the functional
$\vx$. Fixing a functional $\vp$,  we fix all such functionals. We
can therefore call $\vp$  a state of the quantum system.  To avoid
confusion with the term used in the standard quantum mechanics, we
call $\vp$  a elementary state. As a result,  we formulate the
following postulate.

\

{\bf Postulate 3. }

{\it   The result of each individual measurement of observables of
the physical system is determined by the elementary state of this
system.  The elementary state is described by a multilayer
functional $\vp=\{\vx\}$, $\xi \in \Xi $,  defined on $\AAA_+$
whose restriction $\vpx$ to each subalgebra ~$\QQQ$  is a real
homomorphism on the algebra ~$\QQQ$.}

\

We note that we make no additional assumptions about the
properties of the functionals $\vx$.  In particular,  we do not
suppose that

   \beq{1}
 \vx(\A)=\vp_{\xi'}(\A)
\eeq for $\A \in \QQQ\cap\qqq$.  Of course,  equality \rr{1} may
be satisfied for some functionals $\vp$. We say that a functional
$\vp$ is stable on an observable $\A$ if equality \rr{1} holds for
all allowable $\xi$  and~$\xi'$.

Equality \rr{1} seems self-evident.  The possibility of its
violation therefore requires a special comment. A measurement is
the result of an interaction of two systems: the studied physical
system (classical or quantum) and the classical measuring
instrument.  A priori,  the result can depend on both the studied
system and the instrument.  The dependence on the instrument is
parasitic.  To exclude it,  the instrument is calibrated.  The
calibration procedure is essentially as follows.  A test physical
system is taken and subjected to a reproducible measurement. Then
a repeated measurement of this system is performed by the
instrument to be calibrated.  The results of both measurements
should coincide (within the allowable error). Only an instrument
that passes such a test many times deserves the name of a
measuring instrument.

We can thus exclude the dependence of the result on the microstate
of the instrument. But we suppose that according to peculiarities
of the construction,  spatial orientation, or some other
macroscopic properties, every measuring instrument belongs to one
of the types that can be labeled by the parameter $\xi$ ($\xi \in
\Xi$). We assume that the following properties determine whether
the instrument belongs to the type $\xi$.  First, the instrument
is designed for measuring an observable (observables) belonging to
the subalgebra $\QQQ$. Second, for the system in the elementary
state $\vp$, the measurement of an observable $\A \in \QQQ$  gives
the result $A_{\xi}=\vx(\A)$.

It turns out that the dependence on the parameter $\xi$ cannot be
excluded. To prove this statement, it suffices to give at least
one example of a scenario of the interaction of the instrument
with the studied object in which the dependence on $\xi$ is not
excluded by any calibration.

We construct such an example. Let an instrument perform a
reproducible measurement and its influence on the elementary state
of the studied object have the following properties: the
functional $\vx$ remains unchanged, the multilayer functional
$\vp$ becomes stable on the observable $\A$, and if $\B \in \QQQ$
and $\vp$ is stable on $\B$, then it remains stable in the future.
In other respects, $\vp$  changes uncontrollably. An elementary
check shows that instruments with such properties are compatible.
On the other hand, these properties are necessary for the
compatibility of the instruments.

Let $\A \in \QQQ \cap \qqq \quad (\xi \neq \xi')$. We show that
for the instruments of types $\xi$  and  $\xi'$, the possibility
of violation of equality \rr{1} cannot be excluded by any
calibration. To verify equality \rr{1}, we must perform two
measurements for the same physical system: first by the instrument
of type $\xi$ and then by the instrument of type $\xi'$, or vice
versa. It is impossible to perform measurements by two different
instruments absolutely simultaneously.

Let the instrument  of type $\xi$ be used first. We then obtain
the result $A_{\xi}=\vpx$, and the functional $\vp$  is
transformed to the functional $\vp'$. According to the scenario,
this functional should have the property
$\vp'_{\xi'}(\A)=\vp'_\xi(\A)=\vp_\xi(\A)$. The second measurement
(by the instrument of type $\xi'$ ) therefore again gives $A_\xi$.

If we first use the instrument of type $\xi'$ and act identically,
then we obtain the answer $A_{\xi'}$. But it is impossible to find
out if the values $A_\xi$ and $A_{\xi'}$ coincide because we
obtain only one answer in every variant of consecutive
measurements.

We emphasize that the classification of instruments according to
the types $\xi$ is the classification according to the character
of interaction between the instrument  and the studied system. It
is therefore determined not only by the properties of the
instrument but also by the studied system (the algebra \AAA).
Moreover, the instrument can influence the values of observables
that are global for the whole system even if it interacts only
with an isolated part of the system.

The dependence of the measurement result on the type of the
instrument  can be considered a realization and concretization of
Bohr's concepts about the dependence of the result on the general
context of the experiment. At the same time, the proposed variant
of dependence contradicts neither the causality principle nor the
notion about the existence of local reality. But local reality is
not a definite value of every observable for the considered
physical system but the reaction of a measuring instrument  of a
definite type to the elementary state of the system. We can speak
about a definite value of an observable only if the corresponding
elementary state is stable for this observable. The commutative
algebra \AAA{} has only one maximal real commutative subalgebra.
Therefore, in the classical case, all measuring instruments have
the same type, and all elementary states are stable for all
observables.

In view of the above, we can newly interpret the result obtained
by Kochen and Specker in~\cc{ksp}, where a no-go theorem was
proved. The essence of this theorem is that a particle of spin 1
has no internal characteristic (physical reality~\cc{epr})
uniquely determining the values of the squares of the projections
of the spin on three mutually orthogonal directions.

In the approach described in this article, the conditions of the
Kochen-–Specker theorem are not satisfied.  Indeed, the
observables $(\Ss x, \Ss y,\Ss z)$ used in~\cc{ksp} are
compatible. The observables  $(\Ss{x}, \Ss{y'},\Ss{z'})$, where
the directions $x,y'$, and  $z'$  are mutually orthogonal but the
directions $y$ and $z$  are not parallel to the directions $y'$
and $z'$ , are also compatible. The observables $( \Ss y,\Ss z)$
are not compatible with the observables $(\Ss{y'},\Ss{z'})$. The
instruments compatible with the observables $(\Ss x, \Ss y,\Ss z)$
and $(\Ss{x},\Ss{y'},\Ss{z'})$   have different types. These
instruments therefore need not necessarily give the same result
for the measurement of the projection of the spin on the direction
$x$ but this is implicitly assumed in the Kochen-–Specker theorem.

It can be shown~\cc{rud} that the functionals $\vpx$ have the
properties

  \bea{2} &a)& \vx(0)=0; \nn{} &b)& \vx(\I)=1; \nn{}
&c)& \vx(\A^2)\ge 0; \nn{} &d)& \mbox{if } \lll=\vx(\A), \mbox{
then } \lll\in\sss(\A;\QQQ); \nn{} &e)& \mbox{if }
\lll\in\sss(\A;\QQQ), \mbox{ then } \lll=\vx(\A) \mbox{ for some
 } \vx(\A). \eea
  Here, $\sss(\A;\QQQ)$ is the spectrum of the element $\A$
in the subalgebra $\QQQ$. Because the subalgebra~$\QQQ$ is
maximal, the spectrum in the algebra \AAA{} is the same (see,
e.g., \cc{rud}). We recall that, by definition, $\lll \in
\sss(\A;\AAA)$ if and only if the element $ \A-\lll\I $ has no
inverse in the algebra~\AAA.

In the standard quantum mechanics, we must introduce a special
postulate to take the properties of measurements described by
relations~\rr{2} into account. Relations~\rr{2} have a large
constructive potential. In particular, properties (2d) and (2e)
allow constructing the total set of elementary states allowable
for the considered physical system.

To construct a multilayer functional $\vp$, it is clearly
necessary and sufficient to construct all its components $\vx$.
Each functional $\vx$ can be constructed as follows. In the
subalgebra $\QQQ$, we choose an arbitrary system $G(\QQQ)$ of
independent generators. We construct $\vx$ as some map $G(\QQQ)$
onto the set of the points of the spectra of corresponding
elements. On the rest of the elements of the subalgebra $\QQQ$,
the functional $\vx$ is determined by its properties of linearity
and multiplicativity.

The procedure for constructing the functionals $\vx$ is certainly
consistent if the functionals are constructed independently for
different $\xi$. Of course, equality~\rr{1} may be violated in
this case. But we can always construct a multilayer functional
$\vp$ that is stable on all observables belonging to one of the
subalgebras $\QQQ$. For this, it suffices to begin constructing
the functional $\vp$  from this very subalgebra using the
procedure just described. On another subalgebra $\qqq$, we
construct the functional $\vp_{\xi'}$ as follows. Let $\QQQ\cap
\qqq=\QQ_{\xi\xi'}$ and $G(\QQ_{\xi\xi'})$ be independent
generators of the subalgebra $\QQ_{\xi\xi'}$. Let $\bar
G(\QQ_{\xi\xi'})$ be the complement of these generators up to the
set of generators of the subalgebra $\qqq$. If $\A\in
\QQ_{\xi\xi'}$, then we set $\vp_{\xi'}(\A)=\vx(\A)$. If $\A\in
\bar G(\QQ_{\xi\xi'})$,  then we construct $\vp_{\xi'}(\A)$ as a
map $\A$ to one of the points of its spectrum. On the rest of the
elements of the subalgebra $\qqq$, the functional $\vp_{\xi'}$ is
constructed by linearity and multiplicativity.

In quantum measurement, the elementary state cannot be fixed
unambiguously. Indeed, in one experiment, we can measure
observables belonging to the same maximal commutative subalgebra
$\QQQ$ because instruments measuring incompatible observables are
incompatible. As a result, we find only the values of the
functional $\vx$. The rest of the multilayer functional $\vp$
remains undetermined. A repeated measurement using an instrument
of another type will give new information but will uncontrollably
perturb the elementary state that arose after the first
measurement. The information obtained in the first measurement
will therefore become useless.

In this connection, it is convenient to adopt the following
definition.

\

{\sc  Definition 3.}  {\it  Multilayer functionals $\vp$ are said
to be $\vx$-equivalent if they have the same restriction $\vx$ to
the subalgebra $\QQQ$.}

\

In quantum measurement, we can thus find only the equivalence
class to which the studied elementary state belongs. In a
reproducible measurement of observables belonging to the
subalgebra $\QQQ$, we easily recognize the state preparation
procedure of the standard quantum mechanics. In what follows, we
call this state a quantum state and adopt the following
definition.

\

{\sc Definition 4.}  {\it   A quantum state
$\Psi_{\vp\xi}(\,\cdot\,)$ is an equivalence class
$\{\vp\}_{\vp\xi}$ of\quad $\vx$-equivalent elementary states that
are stable for the subalgebra $\QQQ$. }

\

In fact, such a definition of quantum state is convenient only for
systems without identical particles. The point is that the
measuring instrument  cannot distinguish which of the identical
particles it has registered.  It is therefore convenient to
somewhat generalize the notion of the equivalence. We say that a
functional $\vp$   is weakly $\vx$-equivalent to a functional
$\vp'$ if the restriction $\vx$ of the functional $\vp$  to the
subalgebra $\QQQ$ coincides with the restriction $\vp'_{\xi'}$ of
the functional $\vp'$ to the subalgebra $\qqq$. Here, we assume
that the subalgebra $\qqq$  is obtained from the subalgebra $\QQQ$
by replacing the observables of one of the identical particles
with the corresponding observables of another one.

For systems with identical particles, the equivalence should be
replaced with the weak equivalence in the definition of the
quantum state. In what follows, we assume that such a replacement
is made if necessary.

\section{Quantum ensemble and probability theory}

The Kolmogorov probability theory~\cc{kol} is nowadays the most
developed mathematically. It is commonly assumed that a special
quantum probability theory is required for quantum systems. In
this article, we defend the opinion that the classical Kolmogorov
probability theory is also quite sufficient for the quantum case
if we take the peculiarity of quantum measurements into
account~\cc{slav2}.

The Kolmogorov probability theory (see, e.g.,~\cc{kol,nev}) is
based on the notion of the so-called probability space $(\Om,\F,
P)$. The first component $\Om$ is the set (space) of elementary
events. The physical meaning of elementary events is not
explicitly specified, but it is assumed that they are mutually
exclusive and that one and only one elementary event is realized
in each trial. In our case, the role of an elementary event is
played by a elementary state $\vp$. Along with elementary event,
the notion of a "random event" or simply  "event" is introduced.
Each event $F$ is identified with some subset of the set $\Om$.
The event $F$ is assumed to be realized if one of the elementary
events belonging to this set ($\vp\in F$) is realized. It is
assumed that we can find out whether an event is realized or not
in each trial. For elementary events, this requirement is not
imposed. Sets of subsets of the set $\Om$ (including the set $\Om$
itself and the empty set $\emptyset$) are endowed with the
structure of Boolean algebras. The algebraic operations are the
intersection of subsets, the union of subsets,  and the complement
of a subset up to $\Om$. A Boolean algebra that is closed with
respect to countable unions and intersections is called a
$\sss$-algebra.

The second component of a probability space is some $\sss$-algebra
$F$. The set $\Om$, where a fixed $\sss$-algebra $F$ is chosen, is
called a measurable space.

Finally, the third component of a probability space is a
probability measure $P$. This is a map from the algebra $F$ to the
set of real numbers satisfying the conditions (a )$0\leq P(F) \leq
1$ for all $F\in\F$, $P(\Om)=1$ and (b )$P(\sum_j F_j)=\sum_j
P(F_j)$ for every countable family of disjoint subsets $F_j\in
\F$. We note that the probability measure is defined only for the
events belonging to the algebra $F$. The probability is generally
not defined for elementary events.

We now consider the application of the basic principles of
probability theory to the problem of quantum measurements. The
main purpose of a quantum experiment is to find the probability
distributions for some observable quantities. Using a definite
measuring instrument, we can obtain such a distribution for a set
of compatible observables. From the probability theory standpoint,
choosing a certain measuring instrument  corresponds to fixing the
$\sss$-algebra $F$.

We suppose that we conduct a typical quantum experiment. We have
an ensemble of  quantum systems in a definite {\it quantum state}.
For example, we consider particles with the spin 1/2 and the
projection of the spin on the $x$ axis equal to 1/2. We suppose
that we want to investigate the distribution of the projections of
the spin on the directions having the angles $\theta_1$ and
$\theta_2$ to the $x$ axis. The corresponding observations are
incompatible, and we cannot measure both observables in one
experiment. We should therefore conduct two groups of experiments
using different measuring instruments. In our concrete case, the
magnets in the Stern--Gerlach instrument  should have different
spatial orientations.

These two groups of experiments can be described by the respective
probability spaces $(\Om,\F_1,P_1)$ and $(\Om,\F_2,P_2)$. Although
the space of elementary events $\Om$ is the same in both cases,
the probability spaces are different. To endow these spaces with
the measurability property, they are given different
$\sss$-algebras $\F_1$ and  $\F_2$.

Formally and purely mathematically~\cc{nev}, we can construct a
$\sss$-algebra $\F_{12}$  including both the algebras $\F_1$ and
$\F_2$. Such an algebra is called the algebra generated by $\F_1$
and $\F_2$. In addition to the subsets $F^{(1)}_i\in \F_1$ and
$F^{(2)}_j\in \F_2$ of the set $\Om$, it also contains all
intersections and unions of these subsets. But such a
$\sss$-algebra is unacceptable from the physical standpoint.
Indeed, the event $F_{ij}=F^{(1)}_i\cap F^{(2)}_j$ means that the
values of two incompatible observables lie in a strictly  fixed
domain for one quantum object. For a quantum system, it is
impossible in principle to conduct an experiment that could
distinguish such an event. For such an event, the notion
"probability" therefore does not exist at all, i.e., the subset
$F_{ij}$ does not correspond to any probability measure, and the
$\sss$-algebra $\F_{12}$ cannot be used to construct the
probability space. Here, an important peculiar feature of the
application of probability theory to quantum systems is revealed:
not every mathematically possible $\sss$-algebra is physically
allowable.

An element of the measurable space $(\Om,\F)$ thus corresponds in
the experiment to a pair consisting of a quantum object (for
example, in a definite quantum state) and certain type of
measuring instrument  that allows fixing an event of a certain
form. Each such instrument  can separate events corresponding to
some set of compatible observable quantities, i.e., belonging to
the same subalgebra $\QQQ$. If we assume that each measuring
instrument has some type $\xi$, then the $\sss$-algebra $\F$
depends on the parameter $\xi$: $\F=\F_{\xi}$.

In view of the peculiarity of quantum experiments, we should take
care in defining one of the basic notions of probability theory
--— the real random variable. A real random variable is usually
defined as a map from the space $\Om$ of elementary events to the
extended real axis $\RR=[-\infty,+\infty]$. But such a definition
does not take peculiarities of quantum experiments, where the
result may depend on the type of measuring instrument, into
account. We therefore adopt the following definition.

\

{\sc Definition 5.}  {\it   A real random variable is a map from
the measurable space $(\Om,\F_{\xi})$ of elementary events to the
extended real axis.}

\

For an observable $\A$, this means that
$$
\vp\stackrel{\A}{\longrightarrow}A_{\xi}(\vp)\equiv\vx(\A)
\in\RR.$$

We call a set of physical systems of the same type (described by
one algebra \AAA) that are in some quantum state a quantum
ensemble. Experiment gives unambiguous evidence confirming that
such an ensemble has probabilistic properties. We therefore adopt
the following postulate.

\

{\bf Postulate 4.  } {\it   A quantum ensemble can be equipped
with the structure of a probability space.}

\

We consider an ensemble of physical systems in a quantum state
$\Psi_{\vp\eta}(\,\cdot\,)$ $(\eta \in \Xi)$. Correspondingly, we
consider the equivalence class $\{\vp\}_{\vp\eta}$  the space
$\Om(\vp_{\eta})$ of elementary events $\vp$. We suppose that the
value of an observable $\A \in \QQQ$ is measured in an experiment
and an instrument of type $\xi$ is used. Let
$(\Om(\vp_{\eta}),\F_{\xi})$ denote the corresponding measurable
space. Let $P_{\xi}$ be a probability measure on this space, i.e.,
$P_{\xi}(F)$ is the probability of the event $F\in\F_{\xi}$.

We assume that an event $\aA$ is realized in the experiment if the
registered value of the observable $\A$ does not exceed $\aA$. Let
$P_{\xi}(\aA)=P(\vp: \vp_{\xi}(\A)\le\aA)$ denote the probability
of this event. If we know the probabilities $P_{\xi}(F)$, then we
can find the probability $P_{\xi}(\aA)$ using the corresponding
summations and integrations; the distribution $P_{\xi}(\aA)$ is
marginal for the probabilities $P_{\xi}(F)$ (see, e.g.,~\cc{pro}).
The observable $\A$ may belong not only to the subalgebra $\QQQ$
but also to another maximal subalgebra $\qqq$. To find the
probability of the event $\aA$, we can therefore use an instrument
of type $\xi'$. In this case, we might obtain a different value
$P_{\xi'}(\aA)$ for the probability. But experience shows that
probabilities are independent of the measuring instrument used. We
must therefore adopt one more postulate.

\

{\bf Postulate 5.}

{\it   Let an observable $\A\in\QQQ\cap\qqq$. Then the probability
of the event $\aA$ for the system in the quantum state
$\Psi_{\vp\eta}(\,\cdot\,)$ is independent of the type of
instrument used, $P(\vp: \vpx)\le\aA)=
P(\vp:\vp_{\xi'}(\A)\le\aA)$.}

\

Therefore, although $\vp$  is a multilayer functional, we can use
the notation $P(\vp:\vp(\A)\le\aA)$ for the probability of the
event $\aA$.

We suppose that we have an ensemble of quantum systems in a
quantum state $\Psi_{\vp\eta}(\,\cdot\,)$ and conduct a series of
experiments with this ensemble in which an observable $\A$ is
measured. In every real series, we deal with a finite set of
elementary states. In the ideal series, this set can be countable.
Let $\{\vp\}^A_{\vp\eta}$  denote a random countable sample in the
space $\Om(\eta)$ containing all elementary states belonging to
the real series. By the law of large numbers (see,
e.g.,~\cc{nev}), the probability measure $P_{\A}$ in this sample
is uniquely determined by the probabilities
$P(\vp:\vp(\A)\le\aA)$. The probability measure $P_{\A}$
determines the average of the observable $\A$ in the sample
$\{\vp\}^A_{\vp\eta}$:
 \beq{4}
\langle\A\rangle=\int_{\{\vp\}^{A}_{\vp\eta}}\,P_{\A}(d\vp)\,\vp(\A)\equiv
\Psi_{\vp\eta}(\A).
 \eeq

This average is independent of the concrete sample and is
completely determined by the quantum state
$\Psi_{\vp\eta}(\,\cdot\,)$. Formula~\rr{4} defines a functional
(quantum average) on the set $\AAA_+$. This functional is also
denoted by $\Psi_{\vp\eta}(\,\cdot\,)$. The totality of quantum
experiments indicate that we should adopt the following postulate.

\

{\bf  Postulate 6.}

{\it  The functional $\Psi_{\vp\eta}(\,\cdot\,)$ is linear on the
set $\AAA_+$.}

\

\ \\
 This means that $$ \Psi_{\vp\eta}(\A+\B)=\Psi_{\vp\eta}(\A)+\Psi_{\vp\eta}(\B)
\mbox{ also in the case } [\A,\B]\ne0. $$

Each element also in the case of the algebra \AAA{} is uniquely
represented in the form $\R=\A+i\B$, where $\A,\B\in\AAA_+$. The
functional $\Psi_{\vp\eta}(\,\cdot\,)$ can therefore be uniquely
extended to a linear functional on the algebra \AAA:
$\Psi_{\vp\eta}(\R)=\Psi_{\vp\eta}(\A)+i\Psi_{\vp\eta}(\B)$.

We define the seminorm of the element $\R$ by the equality
 \beq{5}
 \|\R\|^2=\sup_{\vx}\vx(\R^*\R)=\rho(\R^*\R),
 \eeq
  where $\rho(\R^*\R)$  is the spectral radius of the element $\R^*\R$
in the algebra \AAA. Such a definition is allowable. First,
$\|\R\|^2\ge 0$ by property (2c). Further, the definition of the
probability measure gives
\beq{6}
\Psi_{\vp\eta}(\R^*\R)=\int_{\{\vp\}^{R^*R}_{\vp\eta}}\,P_{\R^*\R}
(d\vp)\,[\R^*\R](\vp) \le\sup_{\vx}\vx(\R^*\R)=\rho(\R^*\R).
 \eeq
Let $\eta \in \Xi$ be such that $\R^*\R\in\QQ_{\eta}$. Then we
have $\Psi_{\vp\eta}(\R^*\R)=\vp_{\eta}(\R^*\R)$. For such $\eta$,
we obtain
 \beq{7}
\sup_{\vp_{\eta}}\Psi_{\vp\eta}(\R^*\R)=\sup_{\vp_{\eta}}\vp_{\eta}
(\R^*\R)=\rho_{\eta}(\R^*\R),
  \eeq
where $\rho_{\eta}(\R^*\R)$ is the spectral radius in
$\QQ_{\eta}$. Because the subalgebra $\QQ_{\eta}$ is maximal, we
have $\rho_{\eta}(\R^*\R)=\rho(\R^*\R)$. In view of equalities
\rr{5}, \rr{6}, and \rr{7}, it hence follows that
 \beq{8}
 \|\R\|^2=\sup_{\vx}\vx(\R^*\R)=\sup_{\vx}\Psi_{\vp\xi}(\R^*\R).
\eeq
 Because $\Psi_{\vp\xi}(\,\cdot\,)$ is a linear positive functional, the
Cauchy-–Schwarz-–Bunyakovskii inequality holds:

 \beq{9}
|\Psi_{\vp\xi}(\R^*\s)\Psi_{\vp\xi}(\s^*\R)|\le
\Psi_{\vp\xi}(\R^*\R)\Psi_{\vp\xi}(\s^*\s).
 \eeq
It hence follows that the axioms of a seminorm of the element $\R$
are satisfied for $\|\R\|$ (see, e.g., \cc{emch}):
$$\|\R+\s\|\le\|\R\|+\|\s\|, \quad \|\lll\R\|=|\lll|\|\R\|, \quad
\|\R^*\|=\|\R\|.$$

We now consider the set $J$ of the elements $\R$ of the algebra
\AAA{} such that $\|\R\|^2=0$. It follows from inequality~\rr{9}
that $J$ is a two-sided ideal of \AAA. We can therefore form the
quotient algebra $\AAA'=\AAA/J$. In the algebra $\AAA'$, the
relation $\|\R\|^2=0$ implies that $\R=0$. Equality~\rr{5}
therefore defines not a seminorm but a norm in the algebra
$\AAA'$. On the other hand, we can verify that the algebra $\AAA'$
contains the same physical information as \AAA. For this, we
consider two observables $\A$ and $\B$ that either both belong or
both do not belong to each subalgebra $\QQQ$. Let $\A$ and $\B$
satisfy the additional condition $\|\A-\B\|=0$. Then it follows
from equality~\rr{5} that
  \beq{10}
  \vpx=\vx(\B)
  \eeq
for all $\QQQ$  containing these observables. Equality~\rr{10}
means that no experiment can distinguish between these observables
and they can therefore be identified from the phenomenological
standpoint. Passing from the algebra $\AAA$ to the algebra $\AAA'$
allows realizing this identification mathematically. To deal with
an algebra of type $\AAA'$ from the beginning, it is necessary to
adopt the following postulate.

\

{\bf  Postulate 7.}

{\it  If  $\sup_{\vx}|\vx(\A-\B)|=0$, then $\A=\B$.}

\

Postulate 7 has a technical character. At the same time, it
imposes no additional restrictions from the phenomenological
standpoint but only simplifies the mathematical description of
physical systems. In what follows, we assume that Postulate 7 is
satisfied and that equality~\rr{8} therefore defines a norm of the
element $\R$.

The multiplicative properties of the functional $\vx$ imply that
$\vx([\R^*\R]^2)=[\vx(\R^*\R)]^2$. This means that
$\|\R^*\R\|=\|\R\|^2$. Therefore, if we complete the algebra
\AAA{} with respect to the norm $\|\cdot\|$, then \AAA{} becomes a
$C^*$-algebra~\cc{dix}. The algebra of quantum dynamical
quantities can thus be equipped with the structure of a
$C^*$-algebra. In the standard algebraic approach to quantum
theory, this statement is accepted as an initial axiom. Of course,
this is very convenient mathematically. But the necessity of such
an axiom from the phenomenological standpoint remains completely
unclear.

  \section{ $C^*$-algebra and Hilbert space}

A remarkable property of $C^*$-algebras is that every
$C^*$-algebra is isometrically isomorphic to a subalgebra of
linear bounded operators in a suitable Hilbert
space~\cc{emch,dix}. This allows using the customary Hilbert space
formalism below. The connection between $C^*$-algebras and the
Hilbert space is realized by the so-called Gelfand-–Naimark-–Segal
(GNS) construction (see, e.g.,~\cc{emch}). In brief, it is as
follows.

Let \AAA{} be a $C^*$-algebra and $\po$ be a linear positive
functional on this algebra. We assume that two elements
$\R,\,\R'\in\AAA$ are equivalent if the equality
$\po\left(\K^*(\R-\R')\right)=0$ holds for every $\K\in\AAA$. We
let $\Phi(\R)$ denote the equivalence class of the element $\R$
and consider the set $\AAA(\po)$ of all equivalence classes in
\AAA. We convert the set $\AAA(\po)$ into a linear space by
setting $a\Phi(\R)+b\Phi(\s)=\Phi(a\R+b\s)$. We define a scalar
product in $\AAA(\po)$ by the formula
  \beq{11}
\left(\Phi(\R),\Phi(\s)\right)=\po(\R^*\s).
  \eeq
This scalar product determines the norm
$\|\Phi(\R)\|^2=\po(\R^*\R)$ in $\AAA(\po)$. The completion with
respect to this norm converts $\AAA(\po)$ into a Hilbert space.
Each element $\s$ of the algebra \AAA{} is uniquely represented in
this space by the linear operator $\Pi(\s)$ acting by the rule
\beq{12}
 \Pi(\s)\Phi(\R)=\Phi(\s\R).
  \eeq

We consider a subalgebra $\QQQ \quad \xi \in \Xi$. Without loss of
generality, we can consider the elements of this subalgebra as
mutually commuting linear self-adjoint operators in some Hilbert
space. In this case, for each  element $\A \in \QQQ$, we have a
spectral decomposition
 \beq{13}
  \Psi(\A)=\int \Psi(\p(d\lll))\,A(\lll),
  \eeq
where $\Psi$  is an arbitrary bounded positive linear functional
and $\p(d\lll)$ are the projectors of the orthogonal decomposition
of  unity. In what follows, we write the integrals of type \rr{13}
in the form \beq{14}
  \A=\int \p(d\lll)\,A(\lll)
  \eeq
and assume that all integrals (and limits) on the algebra \AAA{}
should be understood in the weak topology. Correspondingly, the
projectors $\p(d\lll)$ satisfy the relation
 \beq{15}
  \I=\int \p(d\lll).
  \eeq
All elements of the subalgebra $\QQQ$  have a common decomposition
of unity. Relations \rr{13}-\rr{15} are purely algebraic and are
independent of the concrete realization of the elements of the
algebra.

Let $\pt$ be a one-dimensional projector. In the operator
representation, this is the projector on a one-dimensional Hilbert
subspace. Such a projector has the properties
 \beq{16}
 \pt^*=\pt, \qquad \pt^2=\pt
  \eeq
and cannot be represented in the form
\beq{17}
 \pt=\sum_{\al}\p_{\al},\quad
 \pt\p_{\al}=\p_{\al}\pt=\p_{\al}\ne\pt.
  \eeq
Properties \rr{16} and \rr{17} can be used as the definition of a
one-dimensional projector as an element of the algebra.

We consider a one-dimensional projector  $\pt \in \QQQ$. Because
\rr{15} is an orthogonal decomposition of unity, we can write
\beq{18}
 \pt\p(d\lll)=\pt\delta(\lll-\tau)\,d\lll.
  \eeq
Let $\{\vp\}_{(\tau)}$ be a set of multilayer functionals such
that $\vx(\pt)=1$. For each observable $\A \in \QQQ$, the value
$\vpx$ is the same for all such functionals. Indeed, using \rr{14}
and \rr{18}, we obtain
  $$ \vpx=\int A(\lll)\Big[\vx(\pt)\delta(\lll-\tau)d\lll +
\vx((1-\pt)\p(d\lll))\Big]=A(\tau).
  $$
The set $\{\vp\}_{(\tau)}$ is therefore a class of
$\vx$-equivalent functionals. We separate its subset
$\{\vp\}_{(\xi\tau)}$ of multilayer functionals that are stable
for all elements of the subalgebra $\QQQ$. This subset determines
a quantum state that is denoted by $\Psi_{\vp\xi\tau}(\,\cdot\,)$
or, briefly, by $\pst(\,\cdot\,)$.

We now consider the GNS construction, where
$\Psi_{\vp\xi\tau}(\,\cdot\,)$ plays the role of the functional
generating the representation. Let $\pht(\pt)$ and $\pht(\I)$ be
the respective equivalence classes of the elements $\pt$ and $\I$.
We verify that $\pht(\pt)=\pht(\I)$. Indeed, by \rr{9}, we have
  $$
\Big|\pst(\R^*(\I-\pt))\Big|^2\le\pst(\R^*\R)\pst(\I-\pt),
  $$
 but $\pst(\I-\pt)=0$ because $\vx(\pt)=1$. By \rr{11} and
 \rr{12}, we have
 \beq{21}
\Big(\pht(\I),\Pi(\s)\pht(\I)\Big)=\pst(\s).
  \eeq
 for every $\s\in\AAA$.

The quantum average over the quantum ensemble $\{\vp\}_{\xi\tau}$
can therefore be represented in the form of the expectation of the
linear operator $\Pi(\s)$ in the state described by the vector
$\pht(\I)=\pht(\pt)$ of the Hilbert space. In the proposed
approach, this allows using the mathematical formalism of the
standard quantum mechanics to the full extent for calculating
quantum averages. But there is an essential difference here
between the proposed approach and the standard quantum mechanics.
In the latter, the relations of type~\rr{21} are postulated (the
Born postulate). This postulate is sufficient for quantum
mechanical calculations, but its necessity is unclear. In
contrast, equality~\rr{21} is a consequence of phenomenologically
necessary postulates in our case.

We now verify that the functional in Postulate 6 exists. We first
show that the functional $\Psi_{\vp\xi\tau}(\,\cdot\,)$ must
satisfy the relation
\beq{22}
\Psi_{\vp\xi\tau}(\s)=\Psi_{\vp\xi\tau}(\pt\s\pt)
  \eeq
for every $\s \in \AAA$.  Indeed, we have
$$
\pst(\pt\s\pt)=\Big(\pht(\I),\Pi(\pt)\Pi(\s)\Pi(\pt)\pht(\I)\Big)=
\Big(\pht(\pt),\Pi(\s)\pht(\pt)\Big)=\pst(\s).
  $$

We now prove the following statement.

\

{\sc Statement.} {\it  If $\A\in\AAA_+$, then
$\A_{\tau}\equiv\pt\A\pt$ has the form $\A_{\tau}=\pt\,\pst(\A)$,
where $\pst(\A)$ is a linear positive functional satisfying the
normalization condition $\pst(\I)=1$.}

\

{\sc Proof.}  We assume that the elements $\A_{\tau}$ and $\pt$
are realized by linear self-adjoint operators. Because
$[\A_{\tau},\pt]=0$, the operators $\A_{\tau}$ and $\pt$ have a
common spectral decomposition of unity. Because the projector
$\pt$ is one-dimensional, the spectral decomposition of
$\A_{\tau}$ should have the form
$\A_{\tau}=\pt\,\pst(\A)+(\I-\pt)\A_{\tau}$. In view of the
relation $\pt\A_{\tau}=\A_{\tau}$, it hence follows that
 \beq{23}
\A_{\tau}= \pt\pst(\A).
  \eeq
The equality
  $$
\pt\pst(\A+\B)=\pt(\A+\B)\pt=\pt\,\pst(\A)+\pt\pst(\B)
  $$
implies the linearity $\pst(\A+\B)=\pst(\A)+\pst(\B)$. By
linearity, the functional $\pst(\A)$ can be extended to the
algebra \AAA: $\pst(\A+i\,\B)=\pst(\A)+i\pst(\B)$, where
$\A,\B\in\AAA_+$.

The positivity of the functional follows from the relation
 $$
\pst(\s^*\s)=\vx(\pt\pst(\s^*\s))= \vx(\pt\s^*\s\pt) \geq 0.
 $$
  Here, we use property (2c). The normalization condition
holds by the equalities
 $$ \pst(\I)=\vx(\pt\pst(\I))=
\vx(\pt\I\pt)=1.
  $$

The functional $\Psi_{\vp\xi\tau}(\,\cdot\,)$ describing the
quantum average therefore has the property required by
Postulate~6. Moreover, it is positive and satisfies the
normalization condition. These are exactly the conditions that
should be satisfied for the functional describing a quantum state.

Because equality \rr{23} is purely algebraic, it holds regardless
of the concrete realization of $\A_{\tau}$ and $\pt$, and the
value of the functional depends only on two factors: on $\pt$
(quantum state) and on $\A$ as an element of the algebra \AAA{}
but not on some particular commutative subalgebra ($\A$ may belong
to several such subalgebras). This means that the functional
$\Psi_{\vp\xi\tau}(\,\cdot\,)$ satisfies Postulate 5.

A functional satisfying relation \rr{22}, where the projector
$\pt$ is one-dimensional, corresponds to a pure quantum state. If
the projector $\pt$ were multidimensional, the state would be
mixed.

\section{Illustration}

We now apply the general arguments to a concrete physical system
—-- the one-dimensional harmonic oscillator.  We are interested in
the Green's functions of this system. Of course, we can pass to
the standard scheme using the Hilbert space via the GNS
construction. But we can propose a more direct approach from the
standpoint of the approach developed here.

We therefore assume that the harmonic oscillator is a physical
system described by the algebra \AAA{} of dynamical quantities
with two noncommuting Hermitian generators $\Q$ and $\PP$
satisfying the commutation relation
 $$
  [\Q,\PP]=i.
  $$
The time evolution in the algebra \AAA{} is governed by the
equation
 $$ \frac{d\,\A(t)}{d\,t}=i\Big[\HH,\A(t)\Big], \qquad
\A(0)=\A,
 $$
where the Hamiltonian has the form $\HH=1/2(\PP^2+\om^2\Q^2)$. The
quantities $\Q,\PP,\A$, and $\HH$ are considered elements of the
abstract algebra \AAA{}. The elements $\Q$, $\PP$ and $\HH$ are
unbounded and therefore do not belong to the $C^*$-algebra. But
their spectral projectors are elements of the $C^*$-algebra, i.e.,
$\Q$, $\PP$ and $\HH$  are elements adjoined to the $C^*$-algebra.
In this case, the algebra \AAA{} can therefore be considered a
$C^*$-algebra completed by adjoined elements.

It is convenient to pass from the Hermitian elements $\Q$, and
$\PP$ to the elements
 $$\am=\frac{1}{\sqrt{2\om}}(\om\Q+i\PP),
\qquad \ap=\frac{1}{\sqrt{2\om}}(\om\Q-i\PP)$$
 with the commutation relation
  \beq{24}
 [\am,\ap]=1
 \eeq
 and the simple time dependence
$$\am(t)=\am\exp(-i\om t), \qquad \ap(t)=\ap\exp(+i\om t).$$

We calculate the generating functional of the Green's functions.
In the standard quantum mechanics, the n-time Green's function is
defined by the formula
 $$ G(t_1,\dots t_n)=\langle
0|T(\Q(t_1)\dots \Q(t_n))|0\rangle,
  $$
   where T is the operator of chronological ordering and
$|0\rangle$ is the quantum ground state.

By the statement proved in the end of the preceding section, the
Green's function in the proposed approach is defined by the
formula
 \beq{25}
  \p_0 T(\Q(t_1)\dots \Q(t_n))\p_0=G(t_1,\dots
t_n)\p_0,
 \eeq
where $\p_0$ is the spectral projector of $\HH$ corresponding to
the minimal energy value. It is easy to see that $\p_0$ can be
represented in the form
 \beq{26}
\p_0 =\lim_{r\to \infty} \exp(-r\ap\am).
 \eeq
As mentioned above, the limit should be understood in the sense of
the weak topology of the $C^*$-algebra.

We first prove the auxiliary statement:
\beq{27}
 \E=\lim_{r_1,r_2\to \infty}
\exp(-r_1\ap\am)(\ap)^k(\am)^l\exp(-r_2\ap\am)=0 \qquad k,l\ge0,
\quad k+l>0.
 \eeq
Let $\Psi$ be a bounded positive linear functional. Then we have .
$$
 \Psi(\E) =\lim_{r_1,r_2\to \infty}\exp(-r_1k-r_2l)\Psi(
(\ap)^k\exp(-r_1\ap\am)\exp(-r_2\ap\am)(\am)^l).
 $$
Here, we use the continuity of the functional $\Psi$ and
commutation relation~\rr{24}. Taking the inequality
$\|\exp(-r\ap\am)\|\le 1$ into account, we further obtain
\begin {eqnarray*}
  |\Psi(\E)|&\leq& \lim_{r_1,r_2\to \infty}\exp(-r_1k-r_2l)
|\Psi( (\ap)^k\exp(-2r_1\ap\am)(\am)^k)|^{1/2}\nn &\times&\quad
|\Psi((\ap)^l\exp(-2r_2\ap\am)(\am)^l)|^{1/2} \nn &\leq&
\lim_{r_1,r_2\to \infty}\exp(-r_1k-r_2l)
|\Psi((\ap)^k(\am)^k)|^{1/2} |\Psi((\ap)^l(\am)^l)|^{1/2}=0.
 \end {eqnarray*}
 This proves equality \rr{27}.

We now verify equality~\rr{26}. In terms of the elements $\ap$ and
$\am$, the Hamiltonian $\HH$ has the form $\HH=\om(\ap\am+1/2)$.
By ~\rr{27}, we have
  $$
\lim_{r_1,r_2\to \infty} \exp(-r_1\ap\am)\HH\exp(-r_2\ap\am)=
\frac{\om}{2}\lim_{r_1,r_2\to \infty} \exp(-(r_1+r_2)\ap\am).
   $$
 This proves equality~\rr{26}.

It follows from formula~\rr{25} that
 \beq{28}
G(t_1,\dots
t_n)\p_0=\lt.\lt(\frac{1}{i}\rt)^n\frac{\delta^n}{\delta
j(t_1)\dots\delta j(t_n)} \p_0 T\exp\lt( i\int^\infty_{-\infty}
dt\,j(t)\Q(t)\rt)\p_0\rt|_{j=0}.
   \eeq
By the Wick theorem ~\cc{bog}), we have
 \bea{29}
&T\exp\lt( i\int^\infty_{-\infty} dt\,j(t)\Q(t)\rt)=\nn& =
\exp\lt( \frac{1}{2i}\int^\infty_{-\infty} dt_1dt_2\,
\frac{\delta}{\delta\Q(t_1)}D^c(t_1-t_2)\frac{\delta}{\delta\Q(t_2)}\rt)
:\exp\lt( i\int^\infty_{-\infty} dt\,j(t)\Q(t)\rt):.
 \eea
Here, $:\;:$ is the normal ordering operation and
$$
D^c(t_1-t_2)=\frac{1}{2\pi}\int dE\,\exp\lt(-i(t_1-t_2)E\rt)
 \frac{1}{\om^2-E^2-i0}. $$

Varying the right-hand side of~\rr{29} with respect to $\Q$ and
taking ~\rr{27} into account, we obtain
 \begin{eqnarray*}
\p_0T\exp\lt( i\int^\infty_{-\infty} dt\,j(t)\Q(t)\rt)\p_0&=&
\exp\lt(- \frac{1}{2i}\int^\infty_{-\infty} dt_1dt_2\,
j(t_1)D^c(t_1-t_2)j(t_2)\rt)\nn \times \p_0:\exp\lt(
i\int^\infty_{-\infty} dt\,j(t)\Q(t)\rt):\p_0 &=& \p_0  \exp\lt(-
\frac{1}{2i}\int^\infty_{-\infty} dt_1dt_2\,
j(t_1)D^c(t_1-t_2)j(t_2)\rt).
 \end{eqnarray*}
Comparison with formula~\rr{28} yields
$$ G(t_1\dots
t_n)=\lt.\lt(\frac{1}{i}\rt)^n\frac{\delta^n Z(j)}
 {\delta j(t_1) \dots \delta j(t_n)}\rt|_{j=0}, $$
  where
   $$ Z(j)=  \exp\lt(\frac{i}{2}\int^\infty_{-\infty} dt_1dt_2\,
j(t_1)D^c(t_1-t_2)j(t_2)\rt) $$
 is the generating functional.

It is well known that considering quantum field systems in the
framework of perturbation theory can be reduced to considering the
multidimensional harmonic oscillator. Therefore, the proposed
method for calculating the generating functional of Green's
functions is immediately generalizable to quantum field models.

\section{Conclusion}

We have attempted to formulate the postulates of quantum mechanics
in a maximally weak form. Almost all postulates are formulated
such that the nonflufillment of each of them not only violates the
mathematical structure of the theory but also leads to a direct
contradiction of experiments. Perhaps only the first postulate,
whose formulation contains unobservable quantities, does not
completely correspond to this assertion.

Despite their weakness, the proposed postulates suffice for
constructing a mathematical formalism including the formalism of
the standard quantum mechanics. At the same time, the proposed
approach clearly shows the applicability domain of this formalism.
It can be used to describe the properties of quantum ensembles. In
exactly this case, "quantum state "is an adequate notion. The
formalism of the standard quantum mechanics is most closely
connected with this notion.

In contrast, "elementary state" is an adequate notion in the case
of a single phenomenon. This notion lies beyond the framework of
the standard quantum mechanics. Its application to such objects as
the "Schr\"odinger cat" or the Universe does not lead to any
paradoxes. In this case, we have no need in such exotic
constructions as the superposition of the live and dead cat or the
multiworld construction of Everett~\cc{ever}. It should be noted
that the problem of quantum paradoxes is completely absent in the
proposed approach.

An important feature of the proposed approach is that it is
equally applicable to both quantum and classical systems although
almost all attention was given to quantum systems in this article.
Nowadays, no serious physicist tries to reduce quantum physics to
the classical one. At the same time, the vast majority believe
that the classical physics is only the limiting case of the
quantum physics, i.e., the classical physics is reducible to the
quantum one. A desire hence arises to quantize all fields
encountered in nature. In the proposed approach, quantum and
classical fields are considered on the same footing, just as
Abelian and non-Abelian fields in the theory of gauge fields. This
opens new possibilities for constructing models in which both
classical and quantum fields are present.

\end {document}